\documentclass[preprint2]{aastex}
\usepackage{epsfig}
\usepackage{color}

\newcommand   {\about} {\mbox{$\sim$}}

\newcommand   {\tim}   {\mbox{$\times$}}

\newcommand   {\arcs}  {\mbox{$^{\prime\prime}$}}

\newcommand   {\kms}   {\mbox{km\,s$^{-1}$}}

\renewcommand {\ga}    {\mbox{\rlap{\hbox{\lower5pt\hbox{$\sim$}}}\hbox{$>$}}}
\renewcommand {\la}    {\mbox{\rlap{\hbox{\lower5pt\hbox{$\sim$}}}\hbox{$<$}}}

\received{2013 June 19}
\accepted{2013 July 17}

\slugcomment{Accepted for publication in The Astrophysical Journal Letters}

\shortauthors{Lis et al.}

\shorttitle{HDO in Comets}

\begin{document}


\title{A \emph{HERSCHEL} STUDY OF D/H IN WATER IN THE JUPITER-FAMILY COMET 45P/HONDA--MRKOS--PAJDU\v{S}\'{A}KOV\'{A} AND PROSPECTS FOR
  D/H MEASUREMENTS WITH CCAT}

\author{D.~C.~Lis$^{1,2}$, N.~Biver$^3$, D.~Bockel\'ee-Morvan$^3$, P.~Hartogh$^4$, E.~A.~Bergin$^5$, G.~A.~Blake$^1$, J.~Crovisier$^3$, M.~de~Val-Borro$^{4,6}$, E.~Jehin$^7$, M.~K\"uppers$^8$, J.~Manfroid$^7$, R.~Moreno$^3$, M.~Rengel$^4$, and S.~Szutowicz$^9$}

\altaffiltext{1}{California Institute of Technology, Pasadena, CA~91125, USA; 
  dcl@caltech.edu, gab@gps.caltech.edu}

\altaffiltext{2}{Visiting professor, LERMA, \'{E}cole Normale Sup\'{e}rieure, Paris, France.}

\altaffiltext{3}{LESIA, Observatoire de Paris, CNRS, UPMC, Universit\'{e}
  Paris-Diderot, Meudon, France;
  nicolas.biver@obspm.fr, dominique.bockelee@obspm.fr, jacques.crovisier@obspm.fr, raphael.moreno@obspm.fr}

\altaffiltext{4}{Max-Planck-Institute f\"ur Sonnensystemforschung, Katlenburg-Lindau, Germany; hartogh@mps.mpg.de, rengel@mps.mpg.de}

\altaffiltext{5}{Astronomy Department, University of Michigan, Ann Arbor, MI, USA; ebergin@umich.edu}

\altaffiltext{6}{Department of Astrophysical Sciences, Princeton University, Princeton, NJ, USA; valborro@princeton.edu}

\altaffiltext{7}{Universite de Liege, Cointe-Ougree, Belgium; jehin@astro.ulg.ac.be, manfroid@astro.ulg.ac.be}

\altaffiltext{8}{European Space Astronomy Center, Madrid, Spain; michael.kueppers@sciops.esa.int}

\altaffiltext{9}{Space Research Centre, Polish Academy of Sciences, Warsaw, Poland; slawka@cbk.waw.pl}

\begin{abstract}
We present \emph{Herschel} observations of water isotopologues in the atmosphere of the Jupiter-family comet 45P/Honda--Mrkos--Pajdu\v{s}\'{a}kov\'{a}. No HDO emission is detected, with a 3$\sigma$ upper limit of $2.0 \times 10^{-4}$ for the D/H ratio. This value is consistent with the earlier \emph{Herschel} measurement in the Jupiter-family comet 103P/Hartley~2. The canonical value of $3 \times 10^{-4}$ measured pre-\emph{Herschel} in a sample of Oort-cloud comets can be excluded at a 4.5$\sigma$ level. The observations presented here further confirm that a diversity of D/H ratios exists in the comet population and emphasize the need for additional measurements with future ground-based facilities, such as CCAT, in the post-\emph{Herschel} era.
\end{abstract}

\keywords{comets: general -- comets: individual (45P/Honda--Mrkos--Pajdu\v{s}\'{a}kov\'{a}) -- molecular processes -- submillimeter: planetary systems}

\section{Introduction}

Studies of the chemical composition and isotopic ratios in cometary materials provide information about the conditions and processes in the outer protosolar nebula. There are two distinct reservoirs of comets in the solar system: the Oort cloud and the Kuiper belt. Until recently it was thought that most short-period, Jupiter-family comets formed in the Kuiper belt region, while most Oort-cloud comets formed nearer the giant planets. However, recent models of the solar system formation suggest a more complex dynamical history (see Section~\ref{discussion}).

Submillimeter observations---combined with simultaneous infrared observations of non-polar species such as CO$_2$, C$_2$H$_2$, and CH$_4$---have provided the most accurate estimates of the composition of the nucleus and isotopic ratios in key species (water, HCN; e.g., \citealt{bockelee00, bockelee05, jehin09}). Deuterium fractionation in interstellar water, as compared to that measured in solar system objects, is of particular interest for understanding the origin and evolution of solar system ices. The Caltech Submillimeter Observatory (CSO) provided the first spectroscopic detection of HDO in a comet \citep{lis97, bockelee98}. Prior to the launch of the \emph{Herschel Space Observatory}\footnote{\emph{Herschel} is an ESA space observatory with science instruments provided by European-led Principal Investigator consortia and with important participation from NASA.}  \citep{pilbratt10}, the D/H ratio in water ratio was measured in only six Oort-cloud comets through observations at submillimeter, infrared and UV wavelengths (see \citealt{jehin09} and references therein). A relatively uniform value of \about 3\tim 10$^{-4}$ was derived, a factor of 12 higher than the protosolar ratio in H$_2$, but significantly lower than values typically measured in the dense interstellar medium (ISM). This suggests that cometary nuclei must include materials thermally processed in the inner nebula that was subsequently transported outward by turbulent diffusion \citep{mousis00, jacquet13}. 

Our view of deuteration in cometary water has changed in light of the recent \emph{Herschel} observations of the Jupiter-family comet 103P/Hartley~2 (103P hereafter) and Oort-cloud comet C/2009 P1 Garradd (\citealt{hartogh11, bockelee12}; see Section~\ref{discussion}). Here we present \emph{Herschel} observations of water and HDO in another Jupiter-family comet, 45P/Honda--Mrkos--Pajdu\v{s}\'{a}kov\'{a} (45P hereafter), and discuss prospects for future submillimeter observations of HDO in cometary atmospheres. Comet 45P returns to the inner solar system every 5.3 yr. It has a small effective radius ($r \sim 0.8$~km; \citealt{lamy05}), but fairly active nucleus (maximum water production rate $\ga 2 \times 10^{28}$~s$^{-1}$), and moves in a very eccentric orbit ($e = 0.83$; $q=0.53$~AU). Its latest apparition (perihelion on 2011 September 28) was especially favorable, with a geocentric distance of only 0.06 AU on 2011 August 15.

\begin{figure}[t] 
\centering 
\includegraphics[width=0.85\columnwidth,angle=0]{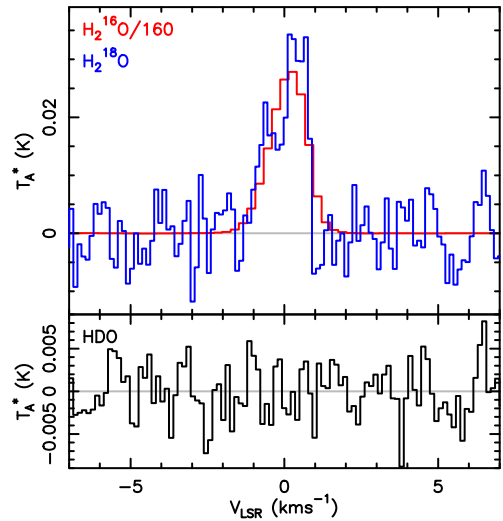} 
\caption{Top: spectra of the $1_{10}$--$1_{01}$ transition of H$_2^{18}$O (blue; HRS) and H$_2^{16}$O line (red, divided by the observed line intensity ratio of 160; WBS) in comet 45P . Bottom: spectrum of the $1_{10}$--$1_{01}$ transition of HDO (HRS).  }
\label{fig:spectra}
\end{figure}

\begin{table}[t] 
    \begin{center}
    \caption{Line Intensities and Production Rates.}\vspace{0.5em}
    \renewcommand{\arraystretch}{1.2}
    \footnotesize
    \begin{tabular}[h]{llcc}
      \hline
      \hline
      Species & Spect & Line~Area  & $Q$ \\
                           &                           &
                           (mK~km\,s$^{-1}$) & (s$^{-1}$) \\
      \hline
       H$_2^{18}$O         & HRS & $42 \pm 2.57$ & $2.28\pm0.14 \times 10^{24}$\\
       HDO                          & HRS & $< 3.93$ & $<3.98 \times 10^{23}$ \\
       H$_2^{16}$O          & WBS & $6630 \pm 5.0$ & $9.10\pm 0.01 \times 10^{26}$\\
      \hline
      \end{tabular}
    \label{table:intensities}
    \end{center}
Notes: Average integrated line intensities for the period 15--21.25~UT on 2011 August 13 were computed over the velocity interval  --1.1 to 0.9 \kms\ for H$_2^{18}$O and HDO, and --2.5 to 2.5 \kms\ for H$_2^{16}$O. Uncertainties quoted for H$_2^{16}$O and H$_2^{18}$O are 1$\sigma$ and the upper limit for HDO is 3$\sigma$. The expansion velocity of 0.75~\kms\ is determined from the observed line profiles.

\end{table}
\section{Observations and Results}\label{observations}

\emph{Herschel} observations of comet 45P\footnote{\emph{Herschel} OBSIDs: 1342227173--1342227194, 1342227197.} were carried out on 2011 August 13, 14.7--22.7 UT, using the Heterodyne Instrument for the Far-Infrared (HIFI; \citealt{degraauw10}), as part of the guaranteed time key program \emph{Water and Related Chemistry in the Solar System} \citep{hartogh09}. The comet was at a heliocentric distance of 1.03~AU and 0.06~AU from \emph{Herschel}. The observing sequence consisted of interleaved frequency-switched (FSW) observations of the $1_{10}$--$1_{01}$ transitions of H$_2^{18}$O and H$_2^{16}$O at 547.676~GHz and 556.939~GHz (simultaneously), and HDO at 509.292~GHz, of 429 and 1633~s duration, respectively. To evaluate the accuracy of the ephemeris and the extent and symmetry of the water outgassing, on-the-fly (OTF) maps of the $1_{10}$--$1_{01}$ transitions of H$_2^{16}$O of 1023~s duration were carried out at the beginning and the end of the observing sequence, followed by a map of the $2_{02}$--$1_{11}$ H$_2^{16}$O line at 987.927~GHz.

\begin{figure*}[t]
\centering
\includegraphics[width=0.9\textwidth,angle=0]{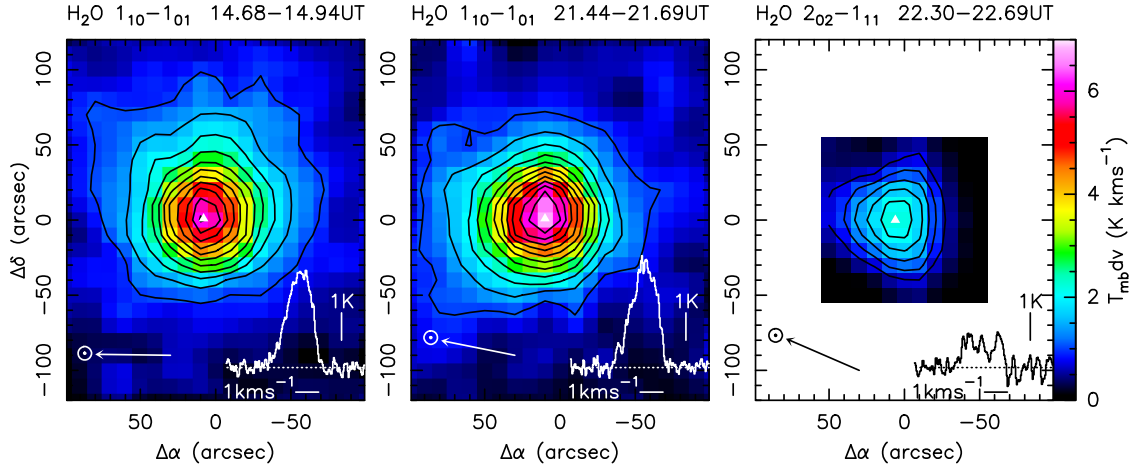}
\caption{HIFI maps of the $1_{10}$--$1_{01}$ and $2_{02}$--$1_{11}$ water emission in comet 45P on 2011 August 13. Contour levels are 1--6 K~\kms, with an interval of 0.5  K~\kms\ for the $1_{10}$--$1_{01}$ transition and 0.667--2 K~\kms, with an interval of 0.333 K~\kms, for the $2_{02}$--$1_{11}$ transition.  Spectra have been interpolated on a regular grid by convolving with a $10^{\prime\prime}$ Gaussian beam for the 557~GHz maps and a $31^{\prime\prime}$ Gaussian beam for the 988~GHz map, to improve the signal-to-noise ratio. The effective angular resolutions are 40$^{\prime\prime}$ and 38$^{\prime\prime}$, respectively (same as the \emph{Herschel} FWHM beam size at 557~GHz). The location of the emission peaks, offset by (9\arcs , 1\arcs ), (10\arcs , 1\arcs ), and (6\arcs , 0\arcs )  with respect to the nucleus, respectively, are marked with white triangles. The maximum uncertainty of the ephemeris is \about 2\arcs\ and the offsets seen in the maps are likely due to anisotropic outgassing in the direction toward the sun (arrows). Spectra shown in the lower-right corner of each panel are averages over a 40\arcs $\times$40\arcs\ region, centered on the peak of the emission. The maps are based on the HRS data. The 557~GHz data correspond to the H instrumental polarization, which generally offers better stability and lower noise level compared to the V polarization. Since the 988~GHz emission is quite weak, the H and V polarization data were averaged to improve the signal-to-noise ratio.
}
\label{fig:maps}
\end{figure*}

The spectra and maps were processed using HIPE \citep{ott10}, pipeline version 10.0.2817. The resulting Level 2 spectra were exported for subsequent data reduction and analysis using the IRAM GILDAS software package.\footnote{http://www.iram.fr/IRAMFR/GILDAS}  All intensities reported here have been corrected for the HIFI main beam efficiency of 75\% and 74\% at 550 and 980~GHz, respectively \citep{roelfsema12}. The full width at half-maximum (FWHM) beam size at the two frequencies is 38\arcs\ and 22\arcs.  The average FSW spectra of H$_2^{18}$O, H$_2^{16}$O, and HDO are shown in Figure~\ref{fig:spectra} and the corresponding integrated line intensities are listed in Table~\ref{table:intensities}.  For the maps, the offsets for each individual OTF spectrum were re-computed based on the UT stamp in the header using the ephemeris JPL\#K112/2 from HORIZONS.\footnote{http://ssd.jpl.nasa.gov/?horizons} The spectra were then interpolated onto a regular grid by convolving with a Gaussian beam of FWHM 10\arcs\ and 31\arcs\ for the 557 and 988 GHz maps, respectively. The resulting maps are shown in Figure~\ref{fig:maps}.

To convert the observed line intensities into production rates (Table~\ref{table:intensities}), we used an excitation model similar to that of Hartogh et al. (2010, 2011), \cite{deval10},  and \cite{bockelee12}. Collisions with electrons use an electron density factor $x_{ne} = 0.2$. A contact surface scaling factor $X_{re}=0.5$ \citep{zakharov07} is used to improve the fit to the maps and the observed 988/557 GHz line ratio. The gas temperature decreases from 150~K near the nucleus to 10~K at an 80~km distance, which provides a good fit to the observed spatial distribution of the water emission (see discussion in \citealt{bockelee12}; in the absence of significant photolytic heating due to the low outgassing rate, we may expect such low temperatures, and an increase of the temperature beyond 6000~km will not affect the observations due to the close proximity of the comet).

\begin{figure*}[t]
\centering
\includegraphics[width=0.9\textwidth,angle=0]{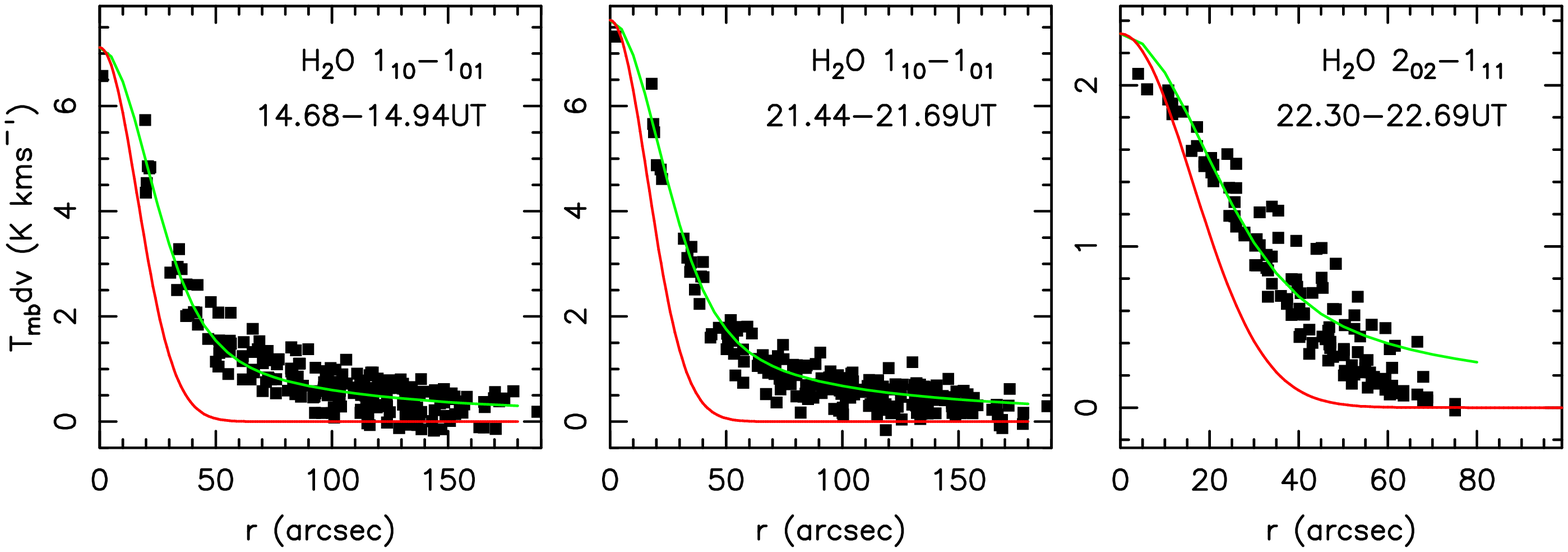}
\caption{Radial profiles of the water emission  in comet 45P, with respect to the peak of the emission, for the three maps shown in Fig. 2.  The $1_{10}$--$1_{01}$ profiles are based on the individual OTF spectra. The $2_{02}$--$1_{11}$ radial profile is based on spectra resampled on a regular grid. Red lines show a 38$^{\prime\prime}$ FWHM Gaussian profile and green lines are predictions of the excitation model discussed in Section~\ref{observations}.
}
\label{fig:profiles}
\end{figure*}
 
The model line intensities as a function of the projected distance from the emission peak, shown as green curves in Figure~\ref{fig:profiles}, are in very good agreement with the observations. The derived water production rate increases with time from $9\times 10^{26}$~s$^{-1}$ at 14.7~UT  to $1.0 \times 10^{27}$~s$^{-1}$ at \about 22~UT. Inspection of the individual FSW spectra indicates that the H$_2^{16}$O production rate during the HDO observing sequence was within 8\% of the mean value reported in Table~\ref{table:intensities}. As a comparison, photometric observations on 2011 July 28 (D.G.~Schleicher, 2013, private communication) give $Q$(OH)$ = 4 \times 10^{26}$~s$^{-1}$ (Haser model) and $Q$(H$_2$O)$ = 5 \times 10^{26}$~s$^{-1}$ (vectorial model), Nan\c{c}ay OH observations give an average production rate of $Q$(OH)$ = 2.3 \pm 0.6 \times 10^{27}$~s$^{-1}$ over 2011 August 19--29 \citep{colom11}.

\begin{figure*}[!ht]
\centering
\includegraphics[trim=70mm 10mm 15mm 80mm, clip, height=150mm,angle=270]{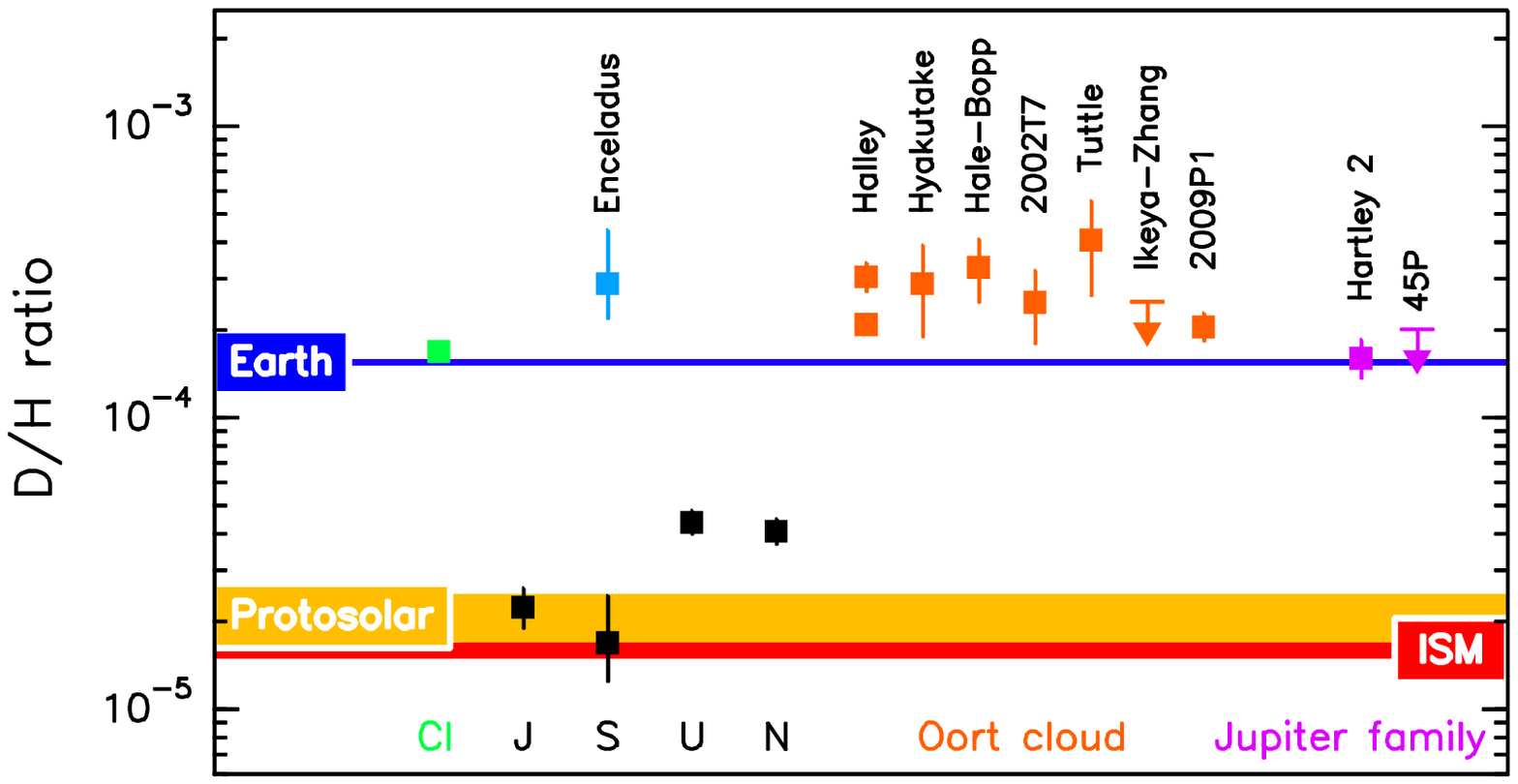}
\caption{Updated version of the figure from \cite{hartogh11} and \cite{bockelee12}, which summarizes D/H values in solar system objects.  Orange symbols are measurements in water in Oort-cloud comets and purple symbols in Jupiter-family comets (current 3$\sigma$ upper limit for 45P). Black symbols are measurements in H$_2$ in atmospheres of giant planets---the Uranus and Neptune values have been revised to take into account the latest \emph{Herschel} measurements \citep{feucht13}. Blue and green symbols are values in water in the plume of Saturn's moon Enceladus and in CI carbonaceous chondrites, respectively. ISM and protosolar values are in H$_2$ and the Earth value is in water. Errorbars are 1$\sigma$. 
}
\label{fig:doverh}
\end{figure*}

We derive a  3$\sigma$ upper limit of D/H$<2.2 \times 10^{-4}$ for water in comet 45P based on H$_2^{16}$O observations. A 3$\sigma$ upper limit for the HDO/H$_2^{18}$O ratio, which is less model dependent, as both lines are optically thin, is 0.175, very close to the Vienna Standard Mean Ocean Water (VSMOW; $0.1554 \pm 0.0001$) and 103P ($0.161 \pm 0.017$) values. For a $^{16}$O/$^{18}$O ratio of 500, this implies D/H $<1.8 \times 10^{-4}$ (3$\sigma$). The \about 25\% discrepancy between the two values is likely due the model assumption of spherical symmetry. We have investigated the consequences of anisotropic outgassing, as evidenced by slight asymmetries in the H$_2^{16}$O maps and spectra (Figure~\ref{fig:maps}), and conclude that a spherically symmetric model may underestimate $Q$(H$_2^{16}$O) and slightly overestimate $Q$(H$_2^{18}$O). The ratio of optically thin H$_2^{18}$O and HDO lines is little affected and we use it in the derivation of D/H. After correcting for the uncertainties in the H$_2^{18}$O line intensity (Table~\ref{table:intensities}) and the $^{16}$O/$^{18}$O isotopic ratio (assumed to be 10\%), our final limit for the D/H in water in comet 45P is $2.0 \times 10^{-4}$ (3$\sigma$). Figure~\ref{fig:doverh} is an updated version of the plot from \cite{hartogh11} and \cite{bockelee12} summarizing current D/H measurements in solar system bodies, which includes the current 45P upper limit and revised \emph{Herschel} values for Uranus and Neptune \citep{feucht13}.

\section{Discussion}\label{discussion}

Studies of protostellar disks around nearby stars, similar to the young Sun, indicate the presence of a radial temperature gradient in the disk, with a temperature at 1~AU from the star in excess of the water thermal desorption temperature. For an optically thin nebula, the ``snow line'' beyond which water ices could exist is at \about 2.7 AU, in the middle of the present-day asteroid belt. This suggests that the Earth accreted ``dry'' and water, and probably also organics, had to be delivered by external sources, such as comets \citep{oro61} or asteroids \citep{alexander12}. In the alternative endogenous scenario, water could survive within the snow line and the Earth formed ``wet'' (e.g., \citealt{saal13}). The D/H isotopic ratio in various solar system reservoirs provides key constraints. 

Pre-\emph{Herschel} measurements in Oort-cloud comets indicated a D/H ratio in water twice the VSMOW value of $(1.558\pm0.001)\times 10^{-4}$. Jupiter-family comets, having formed further away from the Sun, were expected to have even higher D/H values \citep{kavelaars11}. Comets have thus been excluded as the primary source of the Earth's water. \emph{Herschel} provided the first measurement of the D/H ratio in a Jupiter-family comet. A surprisingly low ratio of  ($1.61 \pm 0.24$)~$\times 10^{-4}$ (1$\sigma$), similar to VSMOW, was determined in comet 103P \citep{hartogh11}.  Subsequently, a relatively low D/H ratio of ($2.06 \pm 0.22$)~$\times 10^{-4}$ (1$\sigma$) was also measured in the Oort-cloud comet C/2009 P1 \citep{bockelee12}. 

These observations of comet 45P provide the second measurement of the D/H ratio in water in a Jupiter-family comet. Although HDO was not detected, the derived D/H ratio ($<2.0\times10^{-4}$, 3$\sigma$) is clearly low, possibly lower than the 103P or VSMOW values---the canonical pre-\emph{Herschel} value of $3 \times 10^{-4}$ in Oort-cloud comets can be excluded at a $4.5\sigma$ level. The earlier HIFI observations of comets 103P and C/2009 P1, together with the observations of comet 45P presented here, clearly demonstrate that the high pre-\emph{Herschel} D/H values are not representative of all comets. Based on the current (very small) sample, the D/H ratio in Jupiter-family comets appears to be lower than that in Oort-cloud comets, in disagreement with the classical picture of the comet formation regions and more consistent with models with significant planetesimal interactions and movement \citep{gomes05, morbidelli05}.


The Grand Tack model \citep{walsh11} provides an attractive explanation for the observed morphology of the inner solar system
and the delivery of water to the Earth from the outer asteroid belt. In this model, the Earth's water is accreted during the formation phase of the terrestrial planets and not as a late veneer \citep{morbidelli12}. This, however, appears in contradiction with the recent measurements of a distinct hydrogen isotopic ratio in lunar water, approximately twice that in the Earth's oceans \citep{greenwood11}.
These results seem to suggest that a significant delivery of cometary water to the Earth-Moon system occurred shortly after the Moon-forming impact. The Earth's water would thus be a late addition, resulting from only one, or at most a few collisions with the Earth that missed the Moon \citep{robert11}. However, \cite{saal13} conclude that the lunar magmatic water has an isotopic composition that is indistinguishable from the bulk water in carbonaceous chondrites, and similar to terrestrial water, demonstrating the need for additional studies.

\section{Prospects for HDO Measurements in Comets with CCAT}\label{prospects}

The HIFI instrument provided a unique capability for measurements of the D/H ratio in cometary water. With the end of the \emph{Herschel} cryogenic mission on 2013 April 29, future progress in this field will have to rely on ground-based and airborne (e.g., Stratospheric Observatory for Infrared Astronomy) observations. The capabilities of the Atacama Large Millimeter/submillimeter Array (ALMA) for observations of cometary atmospheres were discussed by \cite{bockelee08}. When the band 8 receivers are completed, ALMA will give access to the 464 GHz ground state HDO transition, first detected in comet C/1996~B2~Hyakutake. Comets are time variable objects that often require scheduling the observations on a short notice, however, and their emission is extended, best recorded in observations with compact ALMA configurations, only available for a fraction of the time. ALMA will thus be best suited for detailed studies of asymmetries in the outgassing of cometary volatiles and its most sensitive use for HDO detection observations in weak comets will be in the autocorrelation mode, with the 50 antennas used as single-dish telescopes.

CCAT\footnote{http://www.ccat.org} is a 25-m diameter submillimeter telescope, to be built near the summit of Cerro Chajnantor in Chile. One of the CCAT first-light instruments currently under development is the CHAI heterodyne array that will cover the 460 GHz and 850 GHz atmospheric windows and offer exceptional capabilities for observations of HDO in cometary atmospheres. One of the main CCAT science objectives will be to significantly increase the sample of comets with accurate D/H measurements and thereby allow statistical studies of the diversity of the isotopic ratio in the Oort-cloud and Jupiter-family comets, as well as possible correlations between the D/H ratio and chemical composition.

CCAT will be able to observe the 509 GHz HDO line\footnote{ In comets at 1~AU from the Sun, the 509~GHz HDO line is typically a few times stronger than the 464~GHz transition accessible to ALMA. The same transition has been utilized in the recent HIFI studies and is currently accessible using the new 460~GHz receiver at the CSO. Based on the existing atmospheric transmission measurements, the expected system temperature at 509~GHz at the CCAT site is the same as at 464~GHz at the ALMA site. The 894~GHz HDO line is even brighter in comets; however, the increase in the system temperature and the decrease in the telescope efficiency at this high frequency compensate for the higher line intensity.} in a typical bright Oort-cloud comet ($Q=10^{29}$ s$^{-1}$, $r_h=0.9$ AU, $\Delta=0.6$ AU) with a signal-to-noise ratio (S/N) of 14 in a 1~hr integration (assuming frequency switching, or switching between two pixels on the CHAI array). A typical Jupiter-family comet ($Q=10^{28}$ s$^{-1}$, $r_h=1$ AU) can be detected with a S/N of 5 out to $\Delta =0.4$ AU in a 10~hr integration. The above computations conservatively assume a low D/H ratio equal to VSMOW---higher S/N detections will be obtained for comets with enhanced D/H ratios. Complementary observations of water in the IR, or UV/radio observations of OH, the photo-dissociation product of water, will be used to obtain accurate water production rates, for comparison with the CCAT HDO measurements. Based on long-term observing statistics, approximately one new, bright Oort-cloud comet per year can be expected. In addition, a Jupiter-family comet bright enough for HDO detection with CCAT will be available every 2--3 yr. Therefore, \about 15 HDO detections can be expected during the first 10 yr of CCAT operations, which would more than double the current sample. 

\acknowledgments 
HIFI has been designed and built by a consortium of institutes and university departments from across Europe, Canada, and the United
States (NASA) under the leadership of SRON, Netherlands Institute for Space Research, Groningen, The Netherlands, and with major contributions from Germany, France and the US. Support for this work was provided by NASA through an award issued by JPL/Caltech. M.dV.B. acknowledges support from NSF (AST-1108686) and NASA (NNX12AH91H) and S.S. from Polish MNiSW (181/N-HSO/2008/0). We thank J.~Giorgini for preparing the ephemeris used in the observations and numerous amateur astronomers who provided astrometric mea  surements on short notice.

\end{document}